\documentclass[aps,preprint]{revtex4}%
\usepackage{amsfonts}
\usepackage{amsmath}
\usepackage{amssymb}
\usepackage{graphicx}%
\providecommand{\U}[1]{\protect \rule{.1in}{.1in}}

\begin{document}
\title[Controlling FFLO momentum through 1D potential]{Controlling the pair momentum of the FFLO state in a 3D Fermi gas through a 1D
periodic potential.}
\author{Jeroen P.A. Devreese$^{1}$}
\email{jeroen.devreese@ua.ac.be}
\author{Michiel Wouters$^{1}$}
\author{Jacques Tempere$^{1,2}$}
\affiliation{$^{1}$TQC (Theory of Quantum systems and Complex systems), Universiteit
Antwerpen, B-2020 Antwerpen, Belgium.}
\affiliation{$^{2}$Lyman Laboratory of Physics, Harvard University, Cambridge, MA 02138, USA.}

\begin{abstract}
The question whether a spin-imbalanced Fermi gas can accommodate the
Fulde-Ferrell-Larkin-Ovchinnikov (FFLO) state has been the subject of intense
study. This state, in which Cooper pairs obtain a nonzero momentum, has
hitherto eluded experimental observation. Recently, we demonstrated that the
FFLO state can be stabilized in a 3D Fermi gas, by adding a 1D periodic
potential. Until now it was assumed that the FFLO wave vector always lies
parallel to this periodic potential (FFLO-P). In this contribution we show
that, surprisingly, the FFLO wave vector can also lie skewed with respect to
the potential (FFLO-S). Starting from the partition sum, the saddle-point free
energy of the system is derived within the path-integral formalism. Minimizing
this free energy allows us to study the different competing ground states of
the system. To qualitatively understand the underlying pairing mechanism, we
visualize the Fermi surfaces of the spin up and spin down particles. From this
visualization, we find that tilting the FFLO wave vector with respect to the
direction of the periodic potential, can result in a larger overlap between
the pairing bands of both spin species. This skewed FFLO state can provide an
additional experimental signature for observing FFLO superfluidity in a 3D
Fermi gas.

\end{abstract}
\date{\today}

\maketitle

\section{Introduction}

Over the last decade, immense progress has been made in the field of the
physics of ultracold atoms. These systems provide a versatile tool for
studying various quantum many-body phenomena \cite{1 - Bloch review,2 -
Greiner Q and A}. One example of the major breakthroughs in this field is the
realization of a molecular Bose-Einstein condensate (BEC) \cite{3 - Greiner,4
- Zwierlein Moleculair BEC} and the observation of pair formation near a
Feshbach resonance \cite{5 - Jochim,6 - Bourdel,7 - Partridge Moleculair BEC,8
- Regal,9 - Zwierlein fermion condensate near Feshbach Reonance} in a strongly
interacting Fermi gas. These experimental discoveries preluded the observation
of superfluidity in a gas of ultracold fermions \cite{10 - Zwierlein fermionic
superfluidity}. The advantage of using ultracold Fermi gases to study
superfluidity is that they allow to explore a wide range of parameter space.
This is due to the tunability of the system parameters, such as the
interaction strength and the spin imbalance. The possibility of adapting the
interaction strength through the use of a Feshbach resonance \cite{11a -
Feshbach resonantie} has allowed to study superfluid pairing in the crossover
from a Bardeen-Cooper-Schrieffer (BCS) state of weakly bound Cooper pairs to a
BEC of tightly bound molecules \cite{11b - BCS-BEC crossover}. Furthermore, in
a Fermi gas composed of a mixture of two hyperfine states (labeled "spin up"
and "spin down"), the ratio between the number of atoms in these two different
states can be controlled with great precision. This last achievement has
provided a unique experimental tool to study the effect of population
imbalance on superfluidity. The first theoretical contribution in this context
was provided by Clogston and Chandrasekhar, who predicted that a first order
transition from a superfluid to a normal interacting Fermi gas would occur at
a critical polarization \cite{12 - Clogston Chandrasekhar}. This quantum phase
transition was indeed observed experimentally \cite{13 - Zwierlein imbalanced
SF,14 - Partridge Hulet imbalanced SF}, along with the fact that when the
system is in the BCS superfluid state, the excess particles are expelled from
this state, leading to phase separation.

The question that has emerged here is whether there exist other, exotic
quantum many body states that allow polarized superfluidity. In 1964, a new
state was introduced independently by Fulde and Ferrell \cite{15 - Fulde
Ferell} and by Larkin and Ovchinnikov \cite{16 - Larkin Ovchinnikov}. They
proposed that a spin-polarized superfluid can be formed by creating Cooper
pairs with finite center-of-mass-momentum. In the last seven years, extensive
theoretical research has been done on this exotic superfluid state \cite{17 -
FFLO artikels}. In 1D and quasi-1D systems, the
Fulde-Ferrell-Larkin-Ovchinnikov (FFLO) state is predicted to be stable
\cite{17b - FFLO 1D and quasi-1D}, and recently the first indirect
experimental evidence for FFLO in a 1D system has been found \cite{17c-Hulet}.
In general, lower dimensionality favors the FFLO state because of Fermi
surface nesting at the FFLO wave vector. For the case of a three dimensional
(3D) Fermi gas, it was found that the FFLO state only occurs on a restricted
area of the BCS-to-BEC-crossover phase diagram \cite{18 - Hu Liu,19 - Sheehy
Radzihovsky}. Up till know, the FFLO state has eluded experimental observation
in a 3D Fermi gas. As an attempt to pave the path towards its experimental
discovery, the idea was proposed to enhance FFLO pairing by using a 3D
periodic optical lattice \cite{20 - Koponen,21 - Trivedi}. Recently, we
proposed to subject the 3D Fermi gas to a one dimensional (1D) periodic
potential as an alternative to achieve this goal \cite{22 - Devreese Klimin
Tempere}. The 1D potential introduces an asymmetry into the system, which
results in an energetically preferred direction for the FFLO wave vector, thus
lowering the energy of this state compared to the normal state and to the BCS state.

In our previous work, it was assumed that the FFLO wave vector lies parallel
to the periodic potential, since this is the energetically preferred
direction. In this paper we show that, contrary to this intuitive expectation,
the wave vector of the FFLO state can also lie skewed with respect to the
direction along which the periodic potential lies. To qualitatively understand
this counterintuitive phenomenon, we present a visualization of the underlying
pairing mechanism by plotting the Fermi surfaces of the spin up and spin down
fermions. This will pinpoint the effect of the periodic potential on the FFLO
pairing mechanism. Our paper is structured as follows. In section
\ref{formalism and calculation} we calculate the free energy of the system
through the use of the path integral formalism \cite{23 - Kleinert,24 - De
Melo}. By choosing an appropriate saddle point, we incorporate the possibility
of the FFLO state in our description, where we allow the FFLO wave vector to
lie in an arbitrary direction. In section \ref{ground states} we minimize the
free energy and discuss the different competing ground states of the system.
Subsequently, in section \ref{phase diagram} we construct the phase diagram as
a function of the total and imbalance chemical potential and in section
\ref{pairing mechanism} we visualize and discuss the changes in the pairing
mechanism of the FFLO state due to the presence of the periodic potential.
Finally in section \ref{conclusion} we draw conclusions.

\section{Path integral treatment\label{formalism and calculation}}

The starting point of our analytic treatment is the partition sum of a 3D
spin-imbalanced Fermi gas, written in path integral form in units $\hbar=2m=1$%
\begin{align}
\mathcal{Z}  &  =\int \mathcal{D}\bar{\psi}\mathcal{D}\psi~\exp \left(
-\sum_{\mathbf{k},n}\sum_{\sigma}\bar{\psi}_{\mathbf{k},\omega_{n},\sigma
}\left[  -i\omega_{n}+\varepsilon \left(  \mathbf{k}_{\bot},k_{z}\right)
-\mu_{\sigma}\right]  \psi_{\mathbf{k},\omega_{n},\sigma}\right. \nonumber \\
&  \left.  -\frac{g}{\beta V}\sum_{\mathbf{k},n}\sum_{\mathbf{q},m}\bar{\psi
}_{\mathbf{k}+\frac{\mathbf{q}}{2},\omega_{n+\Omega_{m}},\uparrow}\bar{\psi
}_{-\mathbf{k}+\frac{\mathbf{q}}{2},-\omega_{n+\Omega_{m}},\downarrow}%
~\psi_{-\mathbf{k}+\frac{\mathbf{q}}{2},-\omega_{n+\Omega_{m}},\downarrow
}~\psi_{\mathbf{k}+\frac{\mathbf{q}}{2},\omega_{n+\Omega_{m}},\uparrow
}\right)  , \label{partition sum}%
\end{align}
with $\beta=1/k_{B}T$ the inverse temperature and $V$ the volume of the
system. In expression (\ref{partition sum}), the fermionic fields are
described by two Grassmann variables $\bar{\psi}$ and $\psi$. Furthermore,
$\mathbf{k=}\left(  \mathbf{k}_{\bot},k_{z}\right)  $ is the single-particle
momentum and $\omega_{n}=\left(  2n+1\right)  \pi/\beta$ and $\Omega_{m}%
=2m\pi/\beta$ represent the fermionic and bosonic Matsubara frequencies
respectively. In the single-particle term, $\varepsilon \left(  \mathbf{k}%
_{\bot},k_{z}\right)  $ is the energy dispersion and the chemical potential
$\mu_{\sigma}$ fixes the number of spin up ($\sigma=\uparrow$) and spin down
($\sigma=\downarrow$) particles. In (\ref{partition sum}), the interaction
between fermions is modeled using a delta-function pseudo potential, given by
$V\left(  \mathbf{r,r}^{\prime}\right)  =g\delta \left(  \mathbf{r-r}^{\prime
}\right)  $, where $g$ is the renormalized interaction strength, which is
given by%
\begin{equation}
\frac{1}{g}=-\frac{1}{8\pi}-\int \frac{d\mathbf{k}}{\left(  2\pi \right)  ^{3}%
}\frac{1}{2\varepsilon \left(  \mathbf{k}_{\bot},k_{z}\right)  },
\end{equation}
where we have set the unit of length to $a_{s}=-1$. We set the scattering
length to a negative value because our aim is to study the FFLO state, which,
to the best of our knowledge, only occurs on the BCS side of the BCS-to-BEC
crossover. In this part of the crossover, the scattering length is always
negative, hence the choice of $a_{s}$.

Our goal is to study the pairing mechanism of the FFLO state under the
influence of a 1D periodic potential. Imposing such a potential on the system
results in a change in the energy dispersion. Here we assume that the
potential is deep enough so that the dispersion is of the following
tight-binding form%
\begin{equation}
\varepsilon \left(  \mathbf{k}_{\bot},k_{z}\right)  =k_{\bot}^{2}+\delta \left[
1-\cos \left(  \frac{\pi k_{z}}{Q_{L}}\right)  \right]  ,
\label{tight-binding dispersion}%
\end{equation}
where $Q_{L}=2\pi/\lambda$ is the wave vector of the periodic potential with
$\lambda$ the wavelength of the periodic potential, and $\delta$ is the
bandwidth. In expression (\ref{tight-binding dispersion}), the periodic
potential is assumed to lie along the z-direction. This convention will be
utilized throughout the rest of this paper. Physically, the tight-binding
approximation corresponds to the situation where the overlap between wave
functions of particles on neighboring sites is taken into account and the
next-to-nearest neighbor hopping is neglected. For a potential depth
$V_{0}\geq4E_{R}$, the tight-binding dispersion
(\ref{tight-binding dispersion}) agrees to within less than five percent with
the exact result \cite{25 - Devreese Wouters Tempere}, where $E_{R}$ is the
recoil energy which in our units is given by $E_{R}=(2\pi/\lambda)^{2}$. In
(\ref{tight-binding dispersion}) the bandwidth $\delta$ is a function of
$E_{R}$ and of the potential depth $V_{0}$. An analytic expression for
$\delta$ can be derived by solving the 1D Mathieu equation in the limit
$V_{0}>>E_{R}$ \cite{26 - Zwerger}. The result is given by
\begin{equation}
\delta=4\left(  \frac{V_{0}^{3}E_{R}}{\pi^{2}}\right)  ^{\frac{1}{4}}%
\exp \left(  -2\sqrt{\frac{V_{0}}{E_{R}}}\right)  .
\end{equation}

To calculate the partition sum (\ref{partition sum}) we use the
Hubbard-Stratonovich transformation, which introduces two auxiliary complex
bosonic fields $\Delta$ and $\bar{\Delta}$ and simultaneously reduces the
fourth order interaction term to two second order interaction terms:
$\bar{\psi}_{\uparrow}\bar{\psi}_{\downarrow}\psi_{\downarrow}\psi_{\uparrow
}\rightarrow \Delta \bar{\psi}_{\uparrow}\bar{\psi}_{\downarrow}+\psi
_{\downarrow}\psi_{\uparrow}\bar{\Delta}$. This transformation takes into
account solely the Bogoliubov channel and is exact as long as there are no
other competing channels that have comparable contributions. For the purpose
of this paper, which focuses on the BCS side of the BCS-BEC crossover and on
zero temperature, the use of Hubbard-Stratonovich is justified. Recently, an
alternative approach to Hubbard-Stratonovich has been introduced \cite{28 -
Kleinert HS} based on Feynman-Kleinert variational perturbation theory
\cite{28a - Feynman Kleinert}. This treatment, however, lies beyond the scope
of the present paper.

Subsequently, the saddle-point approximation is introduced, by which the
bosonic path integral is reduced to the single most contributing term%
\begin{equation}
\int \mathcal{D}\Delta~\exp \left[  -S\left(  \Delta \right)  \right]
\rightarrow \exp \left \{  -S\left[  \Delta_{sp}\left(  \mathbf{q}\right)
\right]  \right \}  ,
\end{equation}
where $\Delta_{sp}$, which in general depends on $\mathbf{q}$, is chosen so
that it minimizes the action $S$. To describe the FFLO state, we take a
specific form for $\Delta_{sp}$ so that the bosonic pairs are allowed to have
a finite center-of-mass momentum $\mathbf{Q}$%
\begin{equation}
\Delta_{sp}\left(  \mathbf{q}\right)  =\sqrt{\beta V}\delta_{\mathbf{q}%
,\mathbf{Q}}\Delta,
\end{equation}
with%
\begin{equation}
\mathbf{Q=}\left(  \mathbf{Q}_{\bot},Q_{z}\right)  .
\label{generalized FFLO pairing mechanism}%
\end{equation}
where the factor $\sqrt{\beta V}$ ensures that $\Delta$ has units of energy.
The two variational parameters $\Delta$ and $\mathbf{Q}$ are interpreted
respectively as the band gap of the system (or, equivalently, as the binding
energy of the bosonic pairs) and the wave vector of the FFLO state. Here, we
allow $\mathbf{Q}$ to have a nonzero perpendicular component $\mathbf{Q}%
_{\bot}$. Since the system exhibits an axial symmetry around the z-axis, it
suffices to determine the magnitude $\left \vert \mathbf{Q}_{\bot}\right \vert $
of this perpendicular component. After applying the saddle-point
approximation, the only remaining path integral is Gaussian and can be
calculated exactly. Finally, the Matsubara summation can be performed
analytically, which results in the following expression for the saddle-point
(sp) free energy $\Omega_{sp}$ that reads, in the zero temperature limit%
\begin{align}
\Omega_{sp}  &  =-\frac{1}{\left(  2\pi \right)  ^{3}}\int_{0}^{\infty}%
dk_{\bot}~k_{\bot}\int_{0}^{2\pi}d\theta \int_{-Q_{L}}^{Q_{L}}dk_{z}\nonumber \\
&  \times \left(  \max \left(  E_{\mathbf{k},\mathbf{Q}},\left \vert
\zeta_{\mathbf{k},\mathbf{Q}}\right \vert \right)  -\xi_{\mathbf{k},\mathbf{Q}%
}-\frac{\left \vert \Delta \right \vert ^{2}}{2\left \{  k_{\bot}^{2}%
+\delta \left[  1-\cos \left(  \frac{\pi k_{z}}{Q_{L}}\right)  \right]
\right \}  }\right)  +\frac{\left \vert \Delta \right \vert ^{2}}{8\pi},
\label{free energy}%
\end{align}
where the following shorthand notations were used%
\begin{align}
\xi_{\mathbf{k},\mathbf{Q}}  &  =k_{\bot}^{2}+\delta \left[  1-\cos \left(
\frac{\pi k_{z}}{Q_{L}}\right)  \cos \left(  \frac{\pi}{2}\frac{Q_{z}}{Q_{L}%
}\right)  \right]  -\left(  \mu-\frac{Q_{\bot}^{2}}{4}\right)
,\label{shorthand 1}\\
\zeta_{\mathbf{k},\mathbf{Q}}  &  =\zeta-\delta \sin \left(  \frac{\pi k_{z}%
}{Q_{L}}\right)  \sin \left(  \frac{\pi}{2}\frac{Q_{z}}{Q_{L}}\right)
+\mathbf{k}_{\bot}\mathbf{.Q}_{\bot},\label{shorthand 2}\\
E_{\mathbf{k},\mathbf{Q}}  &  =\sqrt{\xi_{k,\mathbf{Q}}^{2}+\left \vert
\Delta \right \vert ^{2}}. \label{shorthand 3}%
\end{align}
Here we have introduced $\mu=\left(  \mu_{\uparrow}+\mu_{\downarrow}\right)
/2$ and $\zeta=\left(  \mu_{\uparrow}-\mu_{\downarrow}\right)  /2$ which
represent the total and imbalance chemical potential, respectively. To the
best of our knowledge it is impossible to calculate expression
(\ref{free energy}) completely analytically. However, this expression can be
simplified to a certain degree by integrating out the radial component
$k_{\bot}$. The details of this calculation are given in appendix
\nolinebreak \ref{appendix}.

\section{Results and discussion \label{results and discussion}}

\subsection{The different competing ground states \label{ground states}}

\begin{figure}
[h]
\begin{center}
\includegraphics[
height=8.1939cm,
width=13.3509cm
]%
{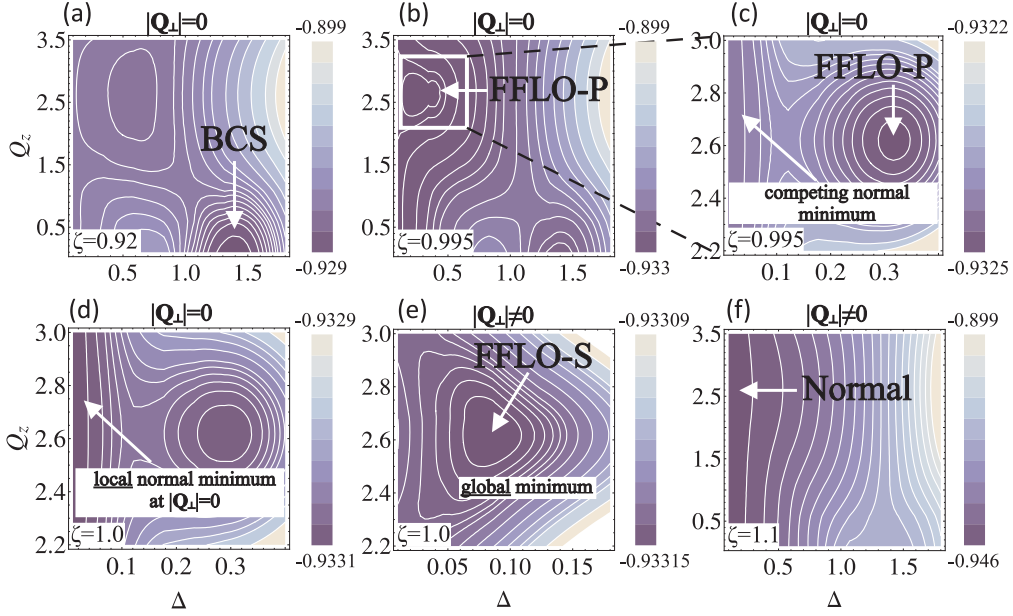}%
\caption{Several contour plots of the free energy as a function of the band
gap $\Delta$ and the z-component of the FFLO wave vector $Q_{z}$. The value of
the perpendicular component $\left \vert \mathbf{Q}_{\bot}\right \vert $ is held
constant in each plot. Darker regions represent lower values of the free
energy. Four different ground states can be identified: (a) a spin-balanced
BCS superfluid $\left(  \Delta \neq0,Q_{z}=0,\left \vert \mathbf{Q}_{\bot
}\right \vert =0\right)  $, (b)/(c) the FFLO-P state $\left(  \Delta \neq
0,Q_{z}\neq0,\left \vert \mathbf{Q}_{\bot}\right \vert =0\right)  $ , (e) the
FFLO-S state $\left(  \Delta \neq0,Q_{z}\neq0,\left \vert \mathbf{Q}_{\bot
}\right \vert \neq0\right)  $, and\ (f) the normal state $\left(
\Delta=0\right)  $. In figures (c) and (d), the emergence of a local normal
minimum that competes with the FFLO-P minimum is shown. There will however be
no first order transition from FFLO-P to normal (c) $\rightarrow$ (d), because
at the transition value of $\zeta$ ($\zeta=1.0$) where the normal minimum lies
lower than the FFLO-P minimum (d), a new global minimum emerges at $\left \vert
\mathbf{Q}_{\bot}\right \vert \neq0$ (e). This means that the system becomes an
FFLO-S state. The behavior shown in figures (c), (d) and (e) only occurs for
$\mu \gtrapprox3.0$. For lower values of $\mu$ the system makes a transition
from BCS over FFLO-P to Normal. In all contour plots we used $\mu=4.44$,
$V_{0}=6$, and $\lambda=1200$ $nm$.}%
\label{figure1.eps}%
\end{center}
\end{figure}

To determine the ground state of the system, we minimize the saddle point free
energy (\ref{free energy}) with respect to the three variational parameters
$\left(  \Delta,Q_{z},\left \vert \mathbf{Q}_{\bot}\right \vert \right)  $, for
given values of the two thermodynamic variables $\mu$ and $\zeta$. The values
of these three variational parameters in a given minimum determine the ground
state of the system, which can be any of the following four states:%
\begin{equation}
\left \{
\begin{array}
[c]{l}%
\text{BCS}\rightarrow \Delta \neq0,Q_{z}=0,\left \vert \mathbf{Q}_{\bot
}\right \vert =0\\
\text{FFLO-P}\rightarrow \Delta \neq0,Q_{z}\neq0,\left \vert \mathbf{Q}_{\bot
}\right \vert =0\\
\text{FFLO-S}\rightarrow \Delta \neq0,Q_{z}\neq0,\left \vert \mathbf{Q}_{\bot
}\right \vert \neq0\\
\text{Normal}\rightarrow \Delta=0
\end{array}
\right.  ,
\end{equation}
where BCS signifies the spin-balanced superfluid state. Here we make a
distinction between the FFLO-P and the FFLO-S state, which respectively denote
that the FFLO wave vector lies parallel ($\left \vert \mathbf{Q}_{\bot
}\right \vert =0$) or skewed ($\left \vert \mathbf{Q}_{\bot}\right \vert \neq0$)
with respect to the direction of the periodic potential. The main question
addressed in this paper is whether the FFLO-S state can be the ground state of
the system, given certain values of $\mu$ and $\zeta$. To show the competition
between the different ground states it is instructive to look at a contour
plot of the free energy as a function of $\Delta$ and $Q_{z}$, as shown in
Fig. \ref{figure1.eps}. The third parameter $\left \vert \mathbf{Q}_{\bot
}\right \vert $ is held constant in a given plot. In Fig. \ref{figure1.eps},
the various plots were made at the same value of $\mu$ ($\mu=4.44$) and for
increasing values of $\zeta$. In Fig. \ref{figure1.eps} (a), for the lowest
value of $\zeta$, the system is in the spin-balanced\ BCS state. This state is
characterized by a nonzero band gap $\Delta$, and by zero momentum
$\mathbf{Q}$ of the fermionic pairs. Figure \ref{figure1.eps} (a) shows that
at finite values of $\zeta$, the Fermi gas can still be spin-balanced. This is
because the nonzero binding energy of the bosonic pairs has to be overcome in
order to break up these pairs. The imbalance chemical potential $\zeta$ can be
interpreted as a Zeeman energy which tries to align the spins in a preferred
direction. When $\zeta$ increases, this magnetic energy increases likewise and
will eventually become larger than the binding energy of the Cooper pairs, at
which point the system will make a transition into a polarized state.%
This transition can be seen in Fig. \ref{figure1.eps} (b), where a second
minimum emerges that starts to compete with the BCS minimum. This new minimum
lies at $\left(  \Delta \neq0,Q_{z}\neq0\right)  $ but still at $\left \vert
\mathbf{Q}_{\bot}\right \vert =0$ and is hence the FFLO-P state, which has a
wave vector parallel to the periodic potential. The transition from BCS to
FFLO-P is of first order, because there is a jump in the value\ of the band
gap $\Delta$. When $\zeta$ increases further, one in principle expects the
FFLO-P state to continuously go over into the normal state, as the band gap
$\Delta$ will gradually decrease to zero when $\zeta$ increases. We indeed
find this behavior at low values of $\mu$ ($\mu \lessapprox3.0$). However, in
the case of Fig. \ref{figure1.eps} (at $\mu=4.44$) the system behaves
differently. When taking a closer look at the FFLO-P minimum in Fig.
\ref{figure1.eps} (b), we see that the normal minimum (at $\Delta=0$) starts
to compete with the FFLO-P minimum, which seems to imply a first order
(FFLO-P)-to-normal transition (see Fig. \ref{figure1.eps} (c) and (d)). This
is not what really happens though, because at the transition value of $\zeta$
($\zeta=1.0$) where the normal minimum lies lower than the FFLO-P minimum
(Fig. \ref{figure1.eps}\ (d)), a new global minimum emerges at $\left \vert
\mathbf{Q}_{\bot}\right \vert \neq0$ (Fig. \ref{figure1.eps} (e)), meaning that
the system has made a transition from the FFLO-P state into the FFLO-S state.
This shows that, at sufficiently high values of the total chemical potential
$\mu$, the system can favor an FFLO state with a wave vector that lies skewed
with respect to the direction of the 1D periodic potential. When $\zeta$ is
increased further, the value of $\Delta$ for the FFLO-S state goes
continuously to zero, and the system makes a transition into a normal
interacting Fermi gas (Fig. \ref{figure1.eps} (f)).

\subsection{The phase diagram as a function of the chemical potentials
\label{phase diagram}}

\begin{figure}
[h]
\begin{center}
\includegraphics[
height=7.856cm,
width=8.9284cm
]%
{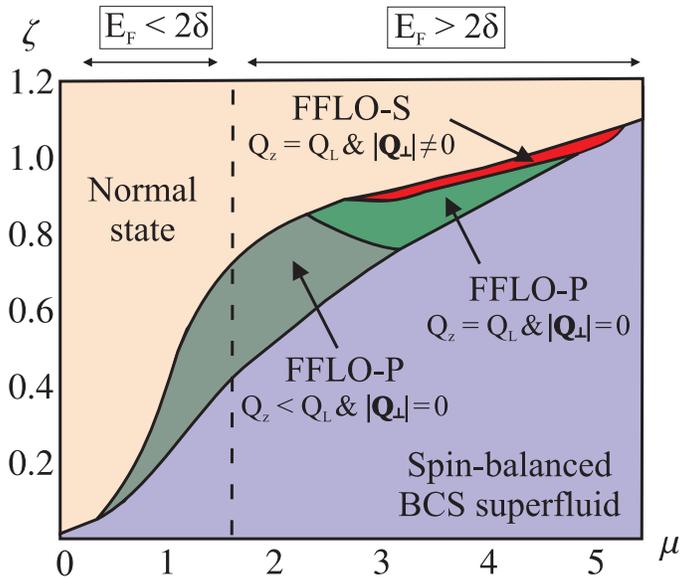}%
\caption{Phase diagram of a spin-imbalanced 3D Fermi gas in the presence of a
1D\ periodic potential, as a function of the total and imbalance chemical
potentials $\mu$ and $\zeta$. The skewed FFLO state (FFLO-S) can be the ground
state of the system but only at higher densities, when the Fermi energy lies
above the energy corresponding to the top of the lowest Bloch band:
$E_{F}>2\delta$.\ The region of the FFLO-P state (wave vector parallel to the
periodic potential) is divided into two parts, where the z-component of the
wave vector of FFLO $Q_{z}$ is either smaller or equal to the laser wave
vector $Q_{L}$. The total FFLO region (FFLO-P + FFLO-S) is significantly
larger than in the known 3D case without periodic potential. In this figure we
used $V_{0}=6$ and $\lambda=1200~nm$ for the depth and wavelength of the
periodic potential respectively. }%
\label{figure2.eps}%
\end{center}
\end{figure}

The examples shown in Fig. \ref{figure1.eps} were only for one particular
value of the total chemical potential $\mu$. When the free energy is minimized
and therefore the ground state is determined for a large set of values for
$\mu$ and $\zeta$, we obtain the phase diagram of the 3D imbalanced Fermi gas
in the presence of a 1D periodic potential at zero temperature, as shown in
Fig. \ref{figure2.eps}.%
This figure shows that the FFLO-S state can indeed be the ground state of the
system. Furthermore, this state only occurs at higher densities, more
specifically when the Fermi energy lies above the energy corresponding to the
top of the first Bloch band: $E_{F}>2\delta$. The FFLO-P region can be
subdivided into two regions. In one region, the z-component $Q_{z}$ of the
wave vector of the FFLO-P state is smaller than the wave vector of the
periodic potential $Q_{L}$. In the other region, this z-component equals the
wave vector of the potential. The FFLO-S state only occurs when $Q_{z}=Q_{L}$.
This observation along with the fact that the FFLO-S state only occurs when
$E_{F}>2\delta$ will be explained in the next subsection, where we illustrate
the FFLO pairing mechanism in momentum space. As we have shown before \cite{22
- Devreese Klimin Tempere}, a striking feature of the phase diagram in Fig.
\ref{figure2.eps} is the significant increase in the area of the FFLO state,
compared to the case of a 3D Fermi gas without periodic potential. In the
latter case, the FFLO area would barely be visible on the scale of Fig.
\ref{figure2.eps}. This is a result of the fact that the periodic potential
lowers the energy for the formation of the FFLO state, relative to the BCS and
the normal state.

\subsection{The influence of the periodic potential on the pairing mechanism
of the FFLO state \label{pairing mechanism}}

\begin{figure}
[ptb]
\begin{center}
\includegraphics[
height=12.9251cm,
width=11.3976cm
]%
{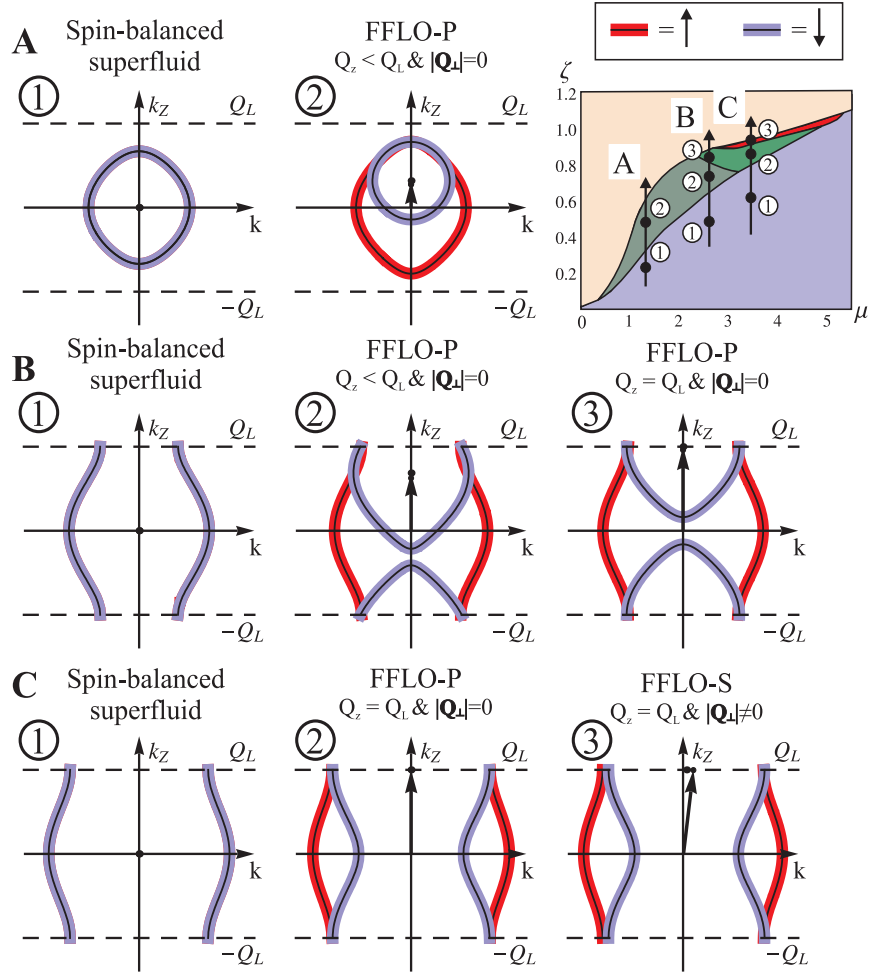}%
\caption{Visualization of the pairing mechanism of the FFLO state in the
presence of a 1D periodic potential for different values of the chemical
potentials. In each picture the Fermi surfaces of up and down spins are shown
in momentum space. Around each Fermi surface a band of thickness $2\Delta$ is
indicated. Only particles that lie in these bands can form superfluid pairs.
The corresponding points in the $\left(  \mu,\zeta \right)  $-phase diagram are
indicated in the inset. In subset A, the Fermi energy lies lower than the
energy corresponding to the top of the lowest Bloch band. In subset B and C,
the Fermi energy lies above the energy of the top of the lowest Bloch band.
When polarization is introduced, the system becomes FFLO-P and the minority
Fermi surface is translated by an amount $Q_{z}$ (A2). When the density or the
polarization increases, the wave vector of FFLO increases likewise in order to
keep both Fermi surfaces aligned (B2). The extreme case is shown in B3 and C2,
where $Q_{z}=Q_{L}$. At even higher density (subset C), the FFLO wave vector
becomes skewed with respect to the direction of the periodic potential (C3).
This costs more energy to translate the Fermi surface, but more pairs can be
formed because the overlap between pairing bands is increased. This lowers the
energy enough for the FFLO-S state to become the ground state.}%
\label{figure3.eps}%
\end{center}
\end{figure}

The occurrence of the FFLO-S state in the $\left(  \mu,\zeta \right)  $-phase
diagram may appear rather counterintuitive at first sight. Indeed, due to the
presence of the periodic potential, it costs less energy to form Cooper pairs
with a momentum that lies parallel to the periodic potential as compared to
Cooper pairs with a momentum that lies skewed with respect to this potential.
To explain why the FFLO-S state can be the ground state of the system, we give
a qualitative description of the pairing mechanism of the FFLO state in the
system under consideration. An insightful visualization of this pairing
mechanism is to plot the Fermi surfaces of up and down particles in momentum
space, as shown in Fig. \ref{figure3.eps}. In this figure, eight different
pictures each show the Fermi surfaces of up and down spins at given values of
the chemical potentials $\mu$ and $\zeta$. Around each Fermi surface, a band
of size $2\Delta$ is indicated. Only particles that lie within these 'pairing
bands' can participate in superfluid pairing. This means that there has to be
an overlap between the pairing bands of up and down spins for superfluid pairs
to exist. Each picture in Fig. \ref{figure3.eps} corresponds to a point on the
phase diagram in Fig. \ref{figure2.eps}. For reasons of clarity, this phase
diagram is reproduced in the inset of figure \ref{figure3.eps}, where the
points corresponding to each of the eight pictures are indicated by small
black filled circles. The different pictures are subdivided in three subsets
A, B and C, which represent distinct physical situations.

Subset A corresponds to the situation where the Fermi energy of the system
lies lower than the energy of the top of the lowest Bloch band $\left(
E_{F}<2\delta \right)  $. In this case, both the Fermi surfaces of up- and
down-particles are approximately spherical and the pairing mechanism is
roughly the same as in the known 3D situation \cite{29 - Chevy}. Picture A1
shows a spin-balanced superfluid, where the Fermi surfaces of both particles
overlap completely since there is no polarization.\ In picture A2, a nonzero
polarization has been introduced by increasing the chemical potential $\zeta$
above a certain critical value. As a consequence, the system has made a
transition into the FFLO-P state, where the magnitude of the wave vector of
the FFLO state is such that it partially re-aligns both Fermi surfaces so that
a maximal number of FFLO-type pairs can be formed. Figure A2 shows the essence
of the FFLO state: granting a nonzero momentum to all pairs requires energy,
but this energy is gained back because more bosonic pairs can be formed. The
presence of the periodic potential favors the FFLO state, since it becomes
energetically more favorable to translate the minority Fermi surface along the
direction of the potential compared to any other direction. This is because in
this direction the energy dispersion (given by expression
(\ref{tight-binding dispersion})) is flatter than a free particle dispersion,
which means that it costs less energy to bridge a gap of given size in
momentum space. In the case of a 3D Fermi gas, the energy dispersion is
quadratic in all dimensions, so adding the periodic potential lowers the
energy for the FFLO state compared to the BCS state and the normal state. This
explains the significant enlargement of the FFLO region in the phase diagram
in Fig. \ref{figure2.eps} compared to the case of a 3D Fermi gas.

Aside from this energy lowering property for the FFLO state, the presence of
the periodic potential has additional effects. This becomes more prominent
when the density of the system is increased and the Fermi energy becomes
larger than the energy corresponding to the top of the lowest Bloch band
$\left(  E_{F}>2\delta \right)  $.%
An example of this situation is shown in subset B in Fig. \ref{figure3.eps}.
Here the value of $\mu$ and hence also the density of the system is larger
compared to subset A. This can for example be seen in picture B1, which
represents a spin-balanced superfluid. In this picture, both Fermi surfaces
are no longer spherical because they have 'hit' the edge of the first
Brillouin zone. When polarization is introduced, the system again becomes of
the FFLO-P type, as shown in pictures B2 and B3. Here it becomes clear why the
wave vector of the FFLO state increases when the polarization is increased. In
order to maximize the overlap between the pairing bands of up and down spins,
the bulge (in the $k_{z}$ direction) in the Fermi surface of the down spins
must align with the waist (in the $k_{z}$ direction) of the Fermi surface of
the up spins. This requires that the Fermi surface of the down spins be
translated increasingly if the polarization is increased. In picture B3, the
polarization is large enough so that the FFLO wave vector becomes equal to the
wave vector of the periodic potential, as indicated by the arrow in this
picture. This is the largest possible value that the FFLO wave vector can
have. Any increase in the magnitude of this vector would simply be equivalent
to an equal decrease in the magnitude of this vector.

In both subsets A and B, the FFLO wave vector always lies parallel to the
direction of the periodic potential. In subset C however, a case is shown
where this no longer applies. Here the density has increased even more
compared to subset B, as can be seen for instance in picture C1. Picture C2
shows another FFLO-P case, where $Q_{z}=Q_{L}$. When the polarization is
increased compared to this case, the majority Fermi surface will widen and the
minority Fermi surface will narrow\ further. This will inevitably result in a
decrease in overlap between the pairing bands of up- and down-spins, thus
raising the energy of the FFLO state compared to the normal state. As can be
seen in picture C3, the FFLO state can solve this problem by forcing its wave
vector to lie skewed with respect to the direction of the periodic potential.
Now it becomes intuitively clear why the FFLO-S state can be the ground state
of the system. Although it costs more energy to add a perpendicular component
to the FFLO wave vector (since the component $Q_{z}$ remains equal to the
laser wave vector $Q_{L}$), more superfluid pairs can be\ formed because the
overlap between the Fermi surfaces of up and down spins is increased, relative
to the case where $\left \vert \mathbf{Q}_{\bot}\right \vert =0$. This is only
true when the density is large enough so that the Fermi energy lies above the
energy of the top of the lowest Bloch band ($E_{F}>2\delta$) and when
$Q_{z}=Q_{L}$ (subset C in \ref{figure3.eps}), which explains why the FFLO-S
state only occurs on that part of the $\left(  \mu,\zeta \right)  $-phase
diagram. One can see the FFLO-S state as a final straw at which the FFLO state
will grasp to maintain its existence in a highly polarized Fermi gas. If the
polarization keeps increasing, the cost in energy will eventually become too
high and the normal state will be energetically more favorable.\newline
Experimentally, the FFLO-S state can provide an additional signature for
observing FFLO superfluidity, in comparison with the FFLO-P state. When
attempting to observe the FFLO-P state, it may be hard to distinguish this
state from the 1D periodic potential. This is because both share the same
Bragg peaks in momentum space when the FFLO state is optimally enhanced
\cite{Acknowledgement Zwierlein}. The advantage of FFLO-S for experimental
detection is that it has an additional momentum component that lies
perpendicular to the momentum component of the periodic potential. Observing
this extra momentum component can provide direct evidence of polarized
superfluidity of the FFLO type.

\section{Conclusion \label{conclusion}}

In this paper, we have studied the effect of a 1D periodic potential on the
pairing mechanism of the FFLO state in a 3D Fermi gas. Starting from the
partition sum of the system, the free energy was derived within the
saddle-point approximation. By choosing a suitable saddle point, we included
the FFLO state in our description, were the FFLO wave vector was allowed to
lie in an arbitrary direction. Minimizing the free energy with respect to the
band gap and the FFLO wave vector allowed to study the different competing
ground states of the system. This subsequently led to the phase diagram as a
function of the total and the imbalance chemical potential. From this phase
diagram we have found that, surprisingly, the wave vector of the FFLO state
can lie skewed (FFLO-S) with respect to the direction along which the periodic
potential lies. Furthermore, we showed that this FFLO-S state only occurs when
the Fermi energy of the system lies above the energy corresponding to the top
of the lowest Bloch band. To gain a deeper understanding of this
counterintuitive phenomenon, we explored the pairing mechanism of the FFLO
state by studying the Fermi surfaces of the spin up and spin down fermions in
momentum space. This visualization showed that the FFLO-S state can be
energetically favorable with respect to the FFLO-P state, because tilting the
FFLO wave vector can result in an increase in the overlap between the pairing
bands of up and down spins. We argued that, experimentally, the FFLO-S state
can provide an additional signature for FFLO superfluidity in a 3D Fermi gas.
This is because this state has an additional perpendicular momentum component,
which allows it to be more easily distinguished from the 1D periodic potential
than the FFLO-P state, since the latter shares the same Bragg peaks with the
potential when the FFLO state is optimally enhanced.

\begin{description}
\item[Acknowledgements] The authors would like to thank Martin Zwierlein and
Hagen Kleinert for helpful suggestions and comments. We wish to thank Randy
Hulet, Carlos S\'{a} de Melo and Sergei Klimin for interesting and stimulating
discussions. One of the authors (JPAD) gratefully acknowledges a Ph. D.
fellowship of the Research Foundation - Flanders (FWO-V). This work was
supported by FWO-V projects G.0356.06, G.0370.09N, G.0180.09N, G.0365.08.
\end{description}

\appendix{}

\section{Analytic treatment of the free energy \label{appendix}}

To the best of our knowledge, it is impossible to calculate all integrals in
expression (\ref{free energy}) for the free energy analytically. It is however
possible to solve the radial part exactly. To accomplish this, we introduce a
cutoff $K_{c}$ in the integral over $k_{\bot}$ and divide expression
(\ref{free energy}) into three parts:%
\begin{equation}
\Omega_{sp}=-\frac{1}{\left(  2\pi \right)  ^{3}}\left(  I_{1}+I_{2}%
+I_{3}\right)  +\frac{\left \vert \Delta \right \vert ^{2}}{8\pi},
\end{equation}
where the three different parts are given by%
\begin{equation}
\left \{
\begin{array}
[c]{l}%
I_{1}=\int_{0}^{2\pi}d\theta \int_{-Q_{L}}^{Q_{L}}dk_{z}\left(  \int_{0}%
^{k_{1}}+...+\int_{k_{N-1}}^{k_{N}}+\int_{k_{N}}^{K_{C}}\right)  dk_{\bot
}~k_{\bot}\max \left(  E_{\mathbf{k},\mathbf{Q}},\left \vert \zeta
_{\mathbf{k},\mathbf{Q}}\right \vert \right) \\
I_{2}=\int_{0}^{2\pi}d\theta \int_{-Q_{L}}^{Q_{L}}dk_{z}\int_{K_{C}}^{\infty
}dk_{\bot}~k_{\bot}\left(  E_{\mathbf{k},\mathbf{Q}}-\xi_{\mathbf{k}%
,\mathbf{Q}}-\frac{\left \vert \Delta \right \vert ^{2}}{2\left \{  k_{\bot}%
^{2}+\delta \left[  1-\cos \left(  \frac{\pi k_{z}}{Q_{L}}\right)  \right]
\right \}  }\right) \\
I_{3}=\int_{0}^{2\pi}d\theta \int_{-Q_{L}}^{Q_{L}}dk_{z}\int_{0}^{K_{C}%
}dk_{\bot}~k_{\bot}\left(  -\xi_{\mathbf{k},\mathbf{Q}}-\frac{\left \vert
\Delta \right \vert ^{2}}{2\left \{  k_{\bot}^{2}+\delta \left[  1-\cos \left(
\frac{\pi k_{z}}{Q_{L}}\right)  \right]  \right \}  }\right)
\end{array}
\right.  , \label{I1 I2 I3}%
\end{equation}
and the shorthand notations (\ref{shorthand 1}) - (\ref{shorthand 3}) are
used. In the first part $I_{1}$, the different values $k_{1},...,k_{N}$
represent the roots of the equation $E_{\mathbf{k},\mathbf{Q}}=$ $\left \vert
\zeta_{\mathbf{k},\mathbf{Q}}\right \vert $, solved with respect to $k_{\bot}$.
These roots are solutions of a fourth order equation and are rather cumbersome
to treat analytically, so it is best to calculate these values numerically.
The main purpose of the division in (\ref{I1 I2 I3}) is to separate the
logarithmic divergence in the integrand in $I_{3}$ $\left(  \text{when }%
k_{z}=0,k_{\bot}=0\right)  $ so that this part can be treated analytically.
The terms in $I_{2}$ have to stay together because they all diverge for
$k_{\bot}\rightarrow \infty$, and they only cancel out when combined. We now
proceed to treat each part analytically to the furthest possible extent.

To calculate the integral over $k_{\bot}$ in $I_{1}$ analytically we consider
two distinct cases. If $E_{\mathbf{k},\mathbf{Q}}>\left \vert \zeta
_{\mathbf{k},\mathbf{Q}}\right \vert $ the result is given by%
\begin{align}
I_{1}  &  =\int_{0}^{2\pi}d\theta \int_{-Q_{L}}^{Q_{L}}dk_{z}\left(  \sum
_{i=1}^{N+2}\right) \nonumber \\
&  \times \left \{  \frac{\Delta^{2}}{4}\log \left(  \frac{\left[  K_{i+1}%
^{2}+A\left(  k_{z}\right)  \right]  +\sqrt{\left[  K_{i+1}^{2}+A\left(
k_{z}\right)  \right]  ^{2}+\Delta^{2}}}{\left[  K_{i}^{2}+A\left(
k_{z}\right)  \right]  +\sqrt{\left[  K_{i}^{2}+A\left(  k_{z}\right)
\right]  ^{2}+\Delta^{2}}}\right)  \right. \nonumber \\
&  +\frac{1}{4}\left(  \left[  K_{i+1}^{2}+A\left(  k_{z}\right)  \right]
\sqrt{\left[  K_{i+1}^{2}+A\left(  k_{z}\right)  \right]  ^{2}+\Delta^{2}%
}\right. \nonumber \\
&  \left.  \left.  -\left(  K_{i}^{2}+A\right)  \sqrt{\left(  K_{i}%
^{2}+A\right)  ^{2}+\Delta^{2}}\right)  \bigskip \right \}  , \label{I1_1}%
\end{align}
where the set of values $K_{i}=\left(  0,k_{1},...,k_{N},K_{c}\right)  $
consists of all roots of equation $E_{\mathbf{k},\mathbf{Q}}=$ $\left \vert
\zeta_{\mathbf{k},\mathbf{Q}}\right \vert $, complemented with zero and the
cutoff $K_{c}$. In the other case when $E_{\mathbf{k},\mathbf{Q}}<\left \vert
\zeta_{\mathbf{k},\mathbf{Q}}\right \vert $ the result is%
\begin{align}
I_{1}  &  =\int_{0}^{2\pi}d\theta \int_{-Q_{L}}^{Q_{L}}dk_{z}\left(  \sum
_{i=1}^{N+2}\right) \nonumber \\
&  \times \left[
\begin{array}
[c]{c}%
\delta_{Q_{\bot}\cos \left(  \theta \right)  ,0}\left(  \frac{\left \vert
B\left(  k_{z}\right)  \right \vert }{2}\left(  K_{i+1}^{2}-K_{i}^{2}\right)
\right) \\
+\left(  \left(  1-\delta_{Q_{\bot}\cos \left(  \theta \right)  ,0}\right)
\left \{  \Theta \left[  Q_{\bot}\cos \left(  \theta \right)  \right]
-\Theta \left[  -Q_{\bot}\cos \left(  \theta \right)  \right]  \right \}
\medskip \right)
\end{array}
\right. \nonumber \\
&  \times \left.  \left(
\begin{array}
[c]{c}%
\left(  \Theta \left[  K_{i}-k_{np}\right]  -\Theta \left[  k_{np}%
-K_{i+1}\right]  \right) \\
\times \left(  \frac{B\left(  k_{z}\right)  }{2}\left(  K_{i+1}^{2}-K_{i}%
^{2}\right)  +\dfrac{Q_{\bot}\cos \left(  \theta \right)  }{3}\left(
K_{i+1}^{3}-K_{i}^{3}\right)  \right) \\
+\Theta \left[  k_{np}-K_{i}\right]  \Theta \left[  K_{i+1}-k_{np}\right] \\
\times \left(  \frac{B\left(  k_{z}\right)  }{2}\left(  K_{i+1}^{2}-2k_{np}%
^{2}+K_{i}^{2}\right)  +\dfrac{Q_{\bot}\cos \left(  \theta \right)  }{3}\left(
K_{i+1}^{3}-2k_{np}^{3}+K_{i}^{3}\right)  \right)
\end{array}
\right)  \right]  , \label{I1_2}%
\end{align}
where the following shorthand notations were used%
\begin{align}
A\left(  k_{z}\right)   &  =\delta \left[  1-\cos \left(  \frac{\pi k_{z}}%
{Q_{L}}\right)  \cos \left(  \frac{\pi}{2}\frac{Q_{z}}{Q_{L}}\right)  \right]
-\left(  \mu-\frac{Q_{\bot}^{2}}{4}\right) \\
B\left(  k_{z}\right)   &  =\zeta-\delta \sin \left(  \frac{\pi k_{z}}{Q_{L}%
}\right)  \sin \left(  \frac{\pi}{2}\frac{Q_{z}}{Q_{L}}\right) \\
k_{np}  &  =\frac{-B\left(  k_{z}\right)  }{Q_{\bot}\cos \left(  \theta \right)
}%
\end{align}
In both expressions (\ref{I1_1}) and (\ref{I1_2}) the set of values
$K_{i}=\left(  0,k_{1},...,k_{N},K_{c}\right)  $ consists of all the roots of
equation $E_{\mathbf{k},\mathbf{Q}}=$ $\left \vert \zeta_{\mathbf{k}%
,\mathbf{Q}}\right \vert $, complemented with zero and with the cutoff $K_{c}$.

Expression $I_{2}$ can also be partly calculated analytically. Here we simply
state the result%
\begin{equation}
I_{2}=\frac{\pi}{2}\int_{-Q_{L}}^{Q_{L}}dk_{z}\left(
\begin{array}
[c]{c}%
\left(  \dfrac{-\Delta^{2}\left[  K_{C}^{2}+A\left(  k_{z}\right)  \right]
}{\left[  K_{C}^{2}+A\left(  k_{z}\right)  \right]  +\sqrt{\left[  K_{C}%
^{2}+A\left(  k_{z}\right)  \right]  ^{2}+\Delta^{2}}}\right)  +\dfrac
{\Delta^{2}}{2}\\
+\Delta^{2}\log \left(  \dfrac{2\left(  K_{C}^{2}+\varepsilon \left(
k_{z}\right)  \right)  }{\left[  K_{C}^{2}+A\left(  k_{z}\right)  \right]
+\sqrt{\left[  K_{C}^{2}+A\left(  k_{z}\right)  \right]  ^{2}+\Delta^{2}}%
}\right)
\end{array}
\right)  ,
\end{equation}
with%
\begin{equation}
\varepsilon \left(  k_{z}\right)  =\delta \left[  1-\cos \left(  \frac{\pi k_{z}%
}{Q_{L}}\right)  \right]  .
\end{equation}
The last part $I_{3}$ contains a logarithmic divergence in the integrand, but
fortunately we found that this expression can be calculated analytically:%
\begin{align}
I_{3}  &  =-\pi K_{C}^{2}Q_{L}\left \{  K_{C}^{2}+2\left[  \delta-\left(
\mu-\frac{Q_{\bot}^{2}}{4}\right)  \right]  \right \} \nonumber \\
&  +Q_{L}\left \vert \Delta \right \vert ^{2}\pi \log \left(  \frac{\delta}{\left(
K_{C}^{2}+\delta \right)  +\sqrt{K_{C}^{4}+2K_{C}^{2}\delta}}\right)  .
\end{align}


\begin{thebibliography}{99}
\bibitem {Acknowledgement Zwierlein}We would like to thank Martin Zwierlein
for pointing out this fact to us.
\end{thebibliography}
\end{document}